\documentstyle[eqsecnum,aps,prb]{revtex} 
\begin{document}
\baselineskip 0.5cm

\vskip 5cm
\title{La substitution induced linear temperature dependence of electrical
resistivity and Kondo behavior in the alloys, Ce$_{2-x}$La$_x$CoSi$_3$}
\vskip 2cm 
\author{Subham Majumdar, M. Mahesh Kumar, R. Mallik, E.V.
Sampathkumaran$^*$} 
\address{Tata Institute of Fundamental Research, Homi
Bhabha Road, Mumbai - 400005, INDIA} \vskip 0.5cm \vskip 1cm \maketitle
\centerline{(Received, \hskip 1cm January 1999 by C.N.R. Rao)}
\begin{abstract} 
The results of electrical resistivity ($\rho$),
heat-capacity (C) and magnetic susceptibility ($\chi$) measurements (above
1.5 K) for the alloys of the type, Ce$_{2-x}$La$_x$CoSi$_3$, are reported. 
We find that the S-shaped temperature dependence of $\rho$ for the
mixed-valent x= 0.0 alloy is not a single-ion property; the most
interesting observation is that, for an intermediate concentration, (x=
0.25), linear temperature dependence of $\rho$ below 30 K is noted, an
observation of relevance to the c urrent trends in the topic of
non-Fermi-liquid behavior. Other observations are: (i) Ce exhibits
single-ionic heavy-fermion behavior with moderately enhanced electronic
contribution to C; and (ii) the strength of the Kondo interaction, as
measured by th e Curie-Weiss parameter gets reduced for larger
substitutions (x $>$ 0.5) of La for Ce.  
\end{abstract} 
\vskip0.5cm

\vskip 0.5cm 
\twocolumn 
\par The
ternary rare-earth (R) compounds of the type, R$_2$TX$_3$ (T= transition
metals, X= Si, Ga), derived from AlB$_2$-type hexagonal structure have
attracted considerable attention in the recent past [1-14]. Perhaps this
class of rare-earth intermetallics
  will be one of the most extensively studied ones next to
ThCr$_2$Si$_2$-type structure. Among these, the compound, Ce$_2$CoSi$_3$,
is of special interest, as the crystallographic analysis reveals ordering
of cobalt atoms within the CoSi$_3$ hexagonal layer and also Ce appears to
be in a mixed-valent state [8]. In this article, we present the results of
electrical resistivity ($\rho$), heat-capacity (C) and magnetic
susceptibility ($\chi$) measurements for the alloys,
Ce$_{2-x}$La$_x$CoSi$_3$, in order
 to probe how the lattice expansion caused by La substitution for Ce
modifies the properties of this compound, paricularly to see whether any
of these alloys exhibit non-Fermi liquid (NFL) behavior, considering
current interest in this direction of resear ch in condensed matter
physics.

The samples, Ce$_{2-x}$La$_x$CoSi$_3$ (x= 0.0, 0.25, 0.5, 1.0, 1.5 and
2.0), have been prepared by melting together stoichiometric amounts of
constituent elements in an arc furnace in an atmosphere of argon. The
samples were annealed at 750$^o$ C for one

 week and then characterized by x-ray diffraction and the diffraction
lines could be indexed on the basis of an AlB$_2$-derived structure. It
appears [8] that the unit cell along the basal plane is twice that of
AlB$_2$-structure, and the lattice constant s derived for this solid
solution (Fig. 1) reveal an expansion of the unit-cell with La
substitution for Ce, as expected. The $\rho$ measurements (1.5-300 K) were
performed by a conventional four-probe method employing a silver-paint for
making electrica l contacts of the leads with the samples. The C
measurements (2-50 K) were performed by a semi-adiabatic heat-pulse
method. The $\chi$ data (2 - 300 K) were taken employing a superconducting
quantum interference device in a magnetic field (H) of 2 kOe.

The results of $\rho$ measurements are shown in Fig. 2. Not much
significance could be attached to the absolute values of $\rho$
considering that the samples are very brittle and porous and hence the
data are normalised to respective 300 K values. The $\rho$ of the parent
Ce compound (x= 0.0) decreases with temperature (T) with a significant,
but continuous fall with T below about 100 K approaching a constant value
at low temperatures. The shape of resistivity versus T plot (usually
called S-shaped) ob served for this composition is typical of mixed-valent
Ce compounds like CeSn$_3$ (Ref. 15). There are qualitative changes in the
shape of $\rho$ versus T plots with the substitution of La for Ce.  It is
interesting to note that small substitutions (x= 0. 25) of La for Ce
transforms low temperature (below 30 K) $\rho$ behavior to a linear T
dependence (NFL behavior), retaining positive T coefficient of $\rho$;
there is a slight deviation from the low temperature linear behavior
around 30 K as the tempe rature is increased, followed by another linear
region in the range 30 - 100 K. For the next higher content of La (x=
0.5), there is a feeble minimum in $\rho$ around 50 K(seen only if the low
temperature data is plotted in an expanded scale), indicative
 of the single-ion Kondo effect, though there is a small drop of $\rho$
below 10 K the origin of which is not clear. For a further addition of La,
a well-defined Kondo minimum in the plot could clearly be seen. Thus, the
S-shaped behavior of the plot of $ \rho$ versus T for x= 0.0 is not a
single-ion property and the data demonstrate Kondo lattice to single-ion
Kondo transformation by La substitution.

The results of C measurements are shown in Fig. 3. It is to be noted that
there is no evidence for the existence of a prominent $\lambda$-type
anomaly for any of the alloys in the entire temperature range of
investigation. This establishes that the paren t Ce compound is
non-magnetic and the possible reduction of the Kondo interaction caused by
La substitution is not sufficient to drive Ce towards magnetic ordering.
For x= 0.0, there is however an extremely small peak around 6 K, which was
observed even in an earlier work [8] and it is not clear whether it is
intrinsic; it is possible that this peak arises from small traces
(estimated to be around 1$\%$, below the detection limit of x-ray
diffraction) of trivalent Ce oxide. The plot of C/T versus T (Fig . 4)
shows an upturn at low temperatures in all cases reaching values above 100
mJ/Ce mol K$^2$ at about 2 K, presumably arising from the electronic
contribution thereby indicating heavy-fermion character of Ce in this
chemical environment. It is not
 clear whether this feature signals NFL nature of the low temperature
state of all these alloys. The value of C/T extrapolated from the linear
regime (12 - 20 K) is also large falling in the range 60 - 80 mJ/Ce mol
K$^2$. We substracted the lattice con tribution to C from the knowledge of
C values of La analogue; we observe that the 4f contribution to C thus
obtained divided by T is independent of T in the range 4 - 20 K, thereby
suggesting that there is no influence of possible high temperature Schot
tky anomalies in this temperature range. Hence we believe that this value
represents true linear term in C.  In short, the results characterize
these alloys as moderate heavy-fermions. 

We could also infer on the dependence of the strength of the Kondo
interaction on composition from the measurement of $\chi$ at the high
temperature range (100-300 K). The plot of inverse $\chi$ versus T is
found to be non-linear in the entire temperatu re range of investigation,
presumably due to large temperature independent contribution (Pauli
susceptibility). Assuming that this contribution is the same as the $\chi$
value of La$_2$CoSi$_3$ at 300 K (4 $\times$ 10$^{-4}$ emu/mol), we
subtract this contribution for each composition. The values thus obtained
(after normalising to Ce concentration) are shown in Fig. 5 in the form of
the plot of inverse $\chi$ versus T. We are now able to clearly see a
linear region in the range 100 - 300 K. The effe ctive moment obtained
from this range turns out to be very close to 2.5 $\mu$$_B$, typical of
trivalent Ce ions for all the compositions and the Curie-Weiss parameter,
$\theta$$_p$ ($\pm$ 10 K) is found to be - 240, -230, -250, - 140 , and
-140 K for x = 0.0, 0.25, 0.5, 1.0 and 1.5 respectively. The trivalency of
Ce and large negative $\theta$$_p$ value for x= 0.0, combined with a large
low temperature electronic contribution to C, establish that the parent Ce
alloy may be classified as a concentrated
 Kondo system.  It is clear from the values of $\theta$$_p$ that the
strength of the Kondo effect at higher temperatures is nearly unaffected
by initial substitutions of La ($<$1.0), whereas the weakening of the
Kondo effect caused by an increase of unit-cell volume is realised only
for higher doping of La. It may be added that there is a deviation from
the high temperature Curie-Weiss behavior at lower temperatures presumably
due to crystal-field effects. There is no ferromagnetic impurity contr
ibution at low temperatures, as the isothermal magnetization is found to
be a linear function of magnetic field.

Summarising, the present results establish that the compound
Ce$_2$CoSi$_3$ is a non-magnetic concentrated Kondo system.  The
transformation from Kondo-lattice to single-ion Kondo behavior is
investigated by the subsitution of La for Ce in this compound. A notable
finding is that, for x= 0.25, we observe linear T-dependence of low
temperature resistivity, a feature which may characterise this composition
as a NFL; it is worthwhile to extend the measurements to low temperatures
(below 1 K) to probe
 whether this NFL behavior persists in such a range as well. Generally
speaking, in non-magnetic Ce-based Kondo lattices, La substitution tends
to drive Ce towards magnetic ordering following the well-known
relationship between unit-cell volume and the Ko ndo effect and
magnetic-coupling strengths [16].  In the present (parent) Ce compound,
the strength of the Kondo interaction is so large (as indicated by
Curie-Weiss parameter) that the lattice expansion caused by La
substitution is not sufficient to ca use magnetic ordering. From the
observation of NFL behavior in resistivity for x= 0.25, we infer that this
alloy can be located across the border separating Fermi-liquids and
magnetic ordering regimes in the Doniach's magnetic phase diagram [16].
Thus, this composition may be near the quantum critical-point and
therefore the magnetic fluctuations presumably lead to non-Fermi-liquid
properties [17].  We also performed $\rho$ measurements with the
application of H and we note that H does not influen ce linear temperature
dependence of resistivity above 4.2 K, similar to the behavior of
U$_{0.9}$Th$_{0.1}$ThBe$_{13}$ (Ref. 18). However, unlike the situation in
this U alloy, the magnetoresistance, MR (till 70 kOe) at 4.2 K is found to
be positive with a magnitude of less than 1$\%$. It is possible that the
positive MR results from Kondo coherence effects in these Ce alloys. It is
of interest to extend MR measurements to very low temperatures ($< $1 K)
in order to see whether Fermi-liquid behavior can b e restored by the
application of magnetic field, an information which will be extremely
valuable to explore possible role of quadrupolar Kondo effect to result in
NFL behavior [18].  We have reported in the past similar quasi-linear T
dependence of resist ivity in a stoichiometric compound, CeIr$_2$Ge$_2$
(Ref. 19). The understanding of such class of alloys in general is a
challenge in condensed matter physics [20]. 
       

{$^*$Electronic address: sampath@tifr.res.in}\\ 
\begin{enumerate}
\item B. Chevalier, P. Lejay, J. Etourneau, and P.
Hagenmuller, Solid State Commun.{\bf 49,} 753 (1984). 

\item P. Kostanidis, P.A. Yakinthos, and E. Gamari-seale, J. Magn. Magn.
Mater. {\bf 87,} 199(1990). 

\item R.E. Gladyshevskii, K. Cenzual, and E. Parthe, J. Alloys Comp.,
{\bf189,} 221(1992). 

\item J.H. Albering, R. P\"ottgen, W. Jeitschko, R.-D. Hoffmann, B.
Chevalier, and Etourneau, J. Alloys Comp., {\bf206,} 133 (1994). 

\item R. P\"ottgen and D.Kaczorowski, J. Alloys Comp., {\bf201,} 157
(1993).  

\item R. P\"ottgen, P. Gravereau, B. Darnet, B. Chevalier, E. Hickey and
J. Etourneau, J. Mater. Chem., {\bf4,} 463 (1994). 

\item A. Raman, Naturwiss., {\bf54,} 560 (1967). 

\item R.A. Gordon, C.J. Warren, M.G. Alexander, F.J. DiSalvo and R.
Pottgen, J. Alloys and Comp. {\bf248}, 24 (1997). 

\item I. Das and E.V. Sampathkumaran, J. Magn. Magn. Mater.
{\bf137,} L239 (1994).

\item R. Mallik, E.V. Sampathkumaran, M. Strecker, G. Wortmann, P.L.
Paulose, and Y. Ueda, J. Magn. Magn. Mater.{\bf185,} L135 (1998). 

\item R. Mallik, E.V. Sampathkumaran, M. Strecker, and G. Wortmann,
Europhys. Lett., {\bf41,} 315 (1998); R. Mallik, E.V. Sampathkumaran, P.L.
Paulose, H. Sugawara, and H. Sato, Pramana - J. Phys, {\bf51,}, 505
(1998). 

\item R. Mallik, E.V. Sampathkumaran, and P.L. Paulose, Solid State
Commun. {\bf106}, 169 (1998). 

\item C. Tien and L. Luo, Solid State Commun., {\bf107}, 295 (1998).

\item J.S. Hwang, K.J. Lin, and C. Tien, Solid State Commun., {\bf100},
169 (1996). 

\item J.M. Lawrence, P.S. Riseborough and R.D. Parks, Rep. Prog. Phys.
{\bf44}, 1 (1981). 

\item S. Doniach, in "Valence Instabilities and Related Narrow Band
Phenomena", edited by R.D. Parks (Plenum, New York, 1977), p. 169;  J.D.
Thompson, Physica B {\bf{223\&224}}, 643 (1996). 

\item M.A. Continentino, Phys. Rep. {\bf{239}}, 179 (1994); Z. Phys. B
{\bf{101}}, 197 (1996). 

\item R.P. Dickey, M.C. de Andrade, J. Hermann, M.B. Maple, F.G. Aliev and
R. Villar, Phys. Rev. B {\bf{56}}, 11 169 (1997). 

\item R. Mallik, E.V. Sampathkumaran, P.L. Paulose, J. Dumschat and G.
Wortmann, Phys. Rev. B {\bf55}, 3627 (1997); R. Mallik, E.V.
Sampathkumaran and P.L. Paulose, Physica B {\bf{230-232}}, 169 (1997). 

\item F.M. Grosche, P. Agarwal, S.R. Julian, N.J. Wilson, R.K.W.
Haselwimmer, S.J.S. Lister, N.D. Mathur, S.V. Carter, S.S. Saxena and G.G.
Lonzarich, Cond-Mat 9812133.  
\end{enumerate} 
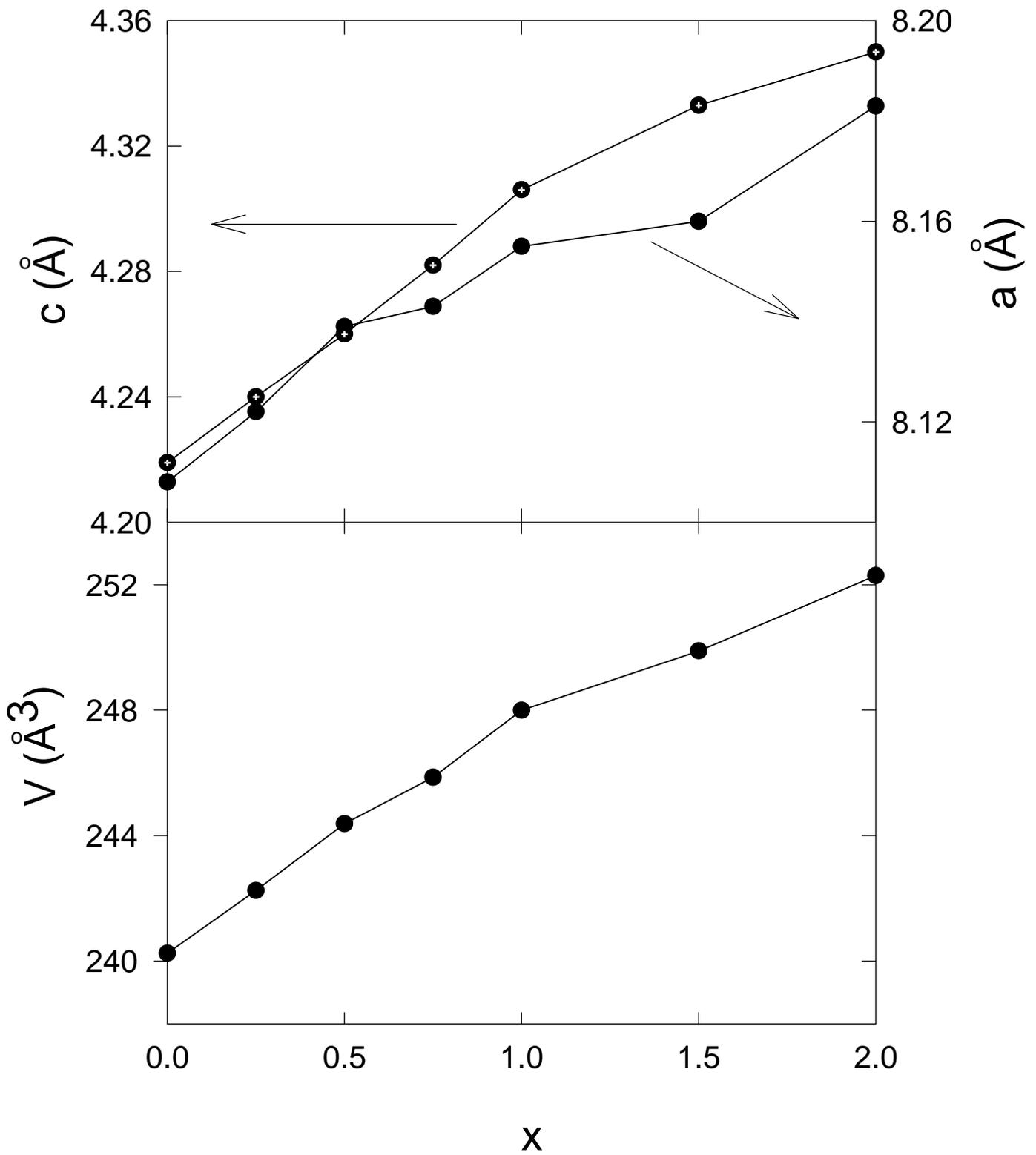
\begin{figure} 
\caption{The lattice constants, a and c, and the unit-cell
volume, V, for the alloys of the series, Ce$_{2-x}$La$_x$CoSi$_3$.}
\end{figure} 
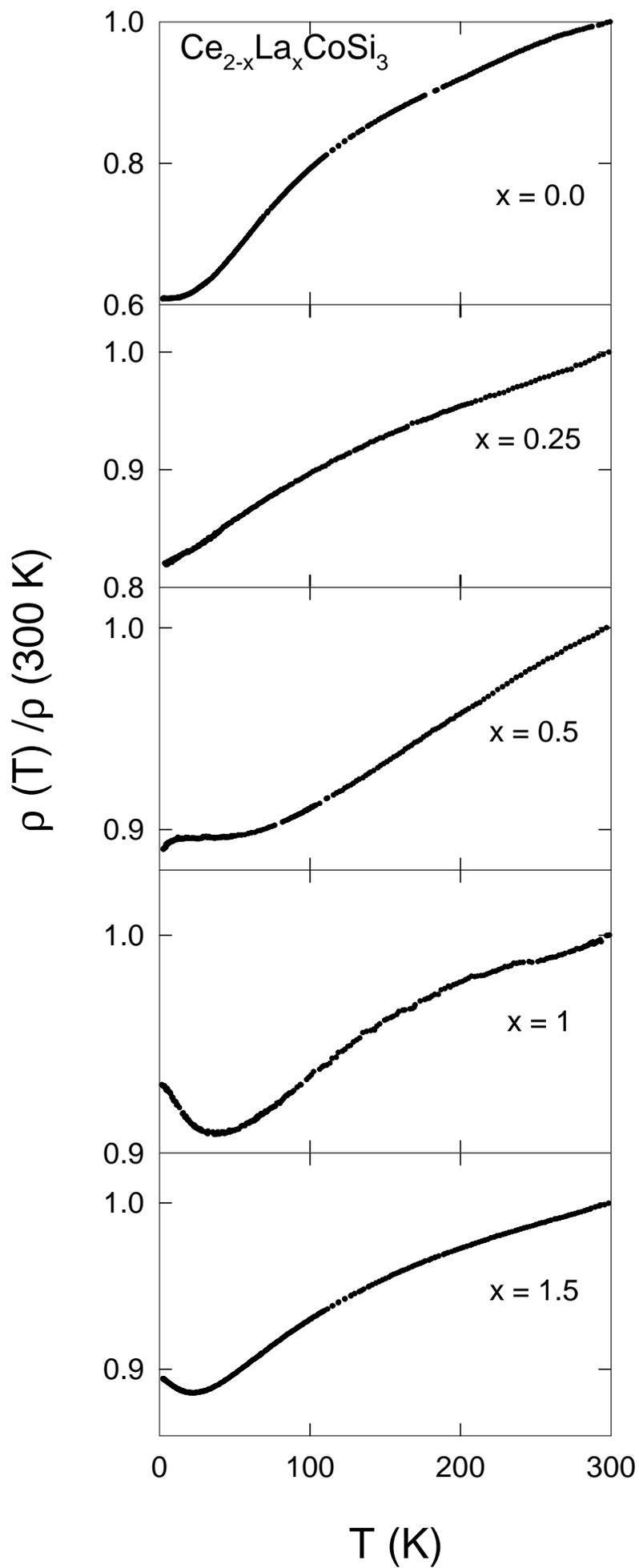
\begin{figure} 
\caption{The electrical resistivity as a
function of temperature for the alloys, Ce$_{2-x}$La$_x$CoSi$_3$,
normalised to respective 300 K values.} 
\end{figure} 
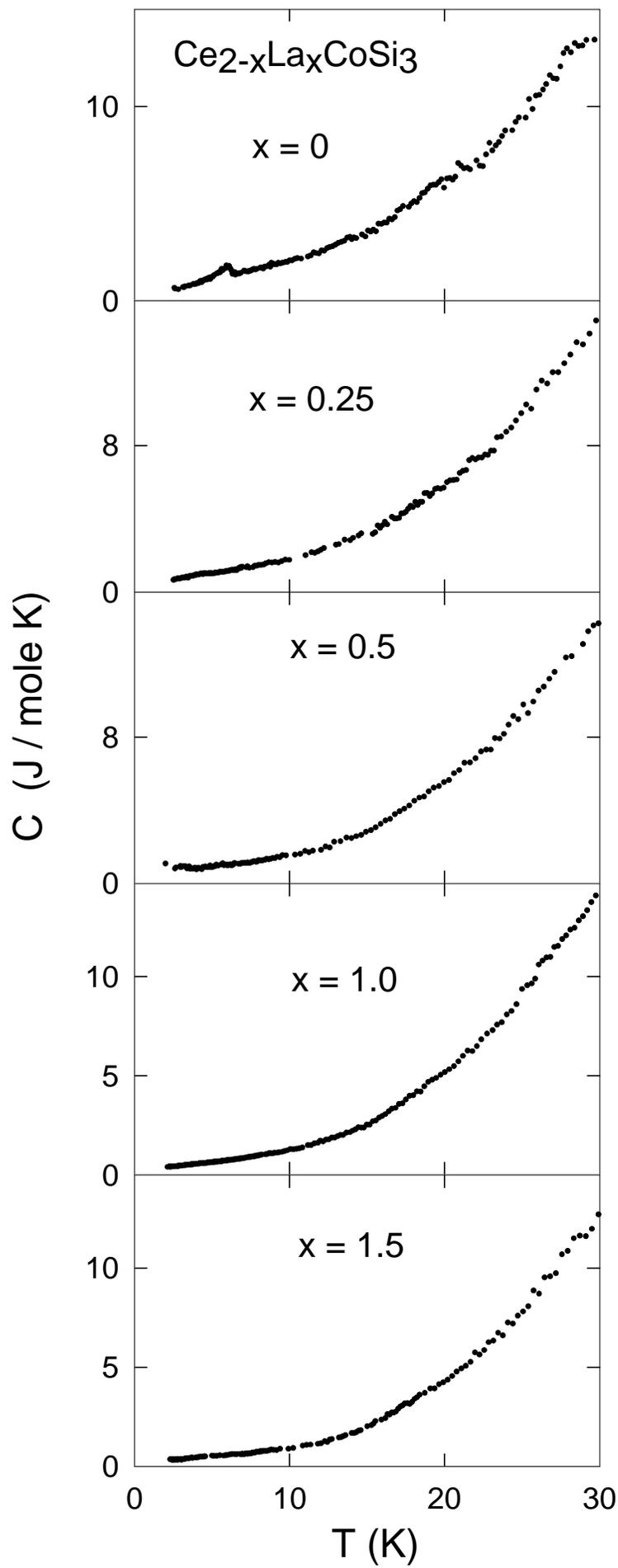
\begin{figure}
\caption{The heat capacity as a function of temperature for the alloys,
Ce$_{2-x}$La$_x$CoSi$_3$.} 
\end{figure} 
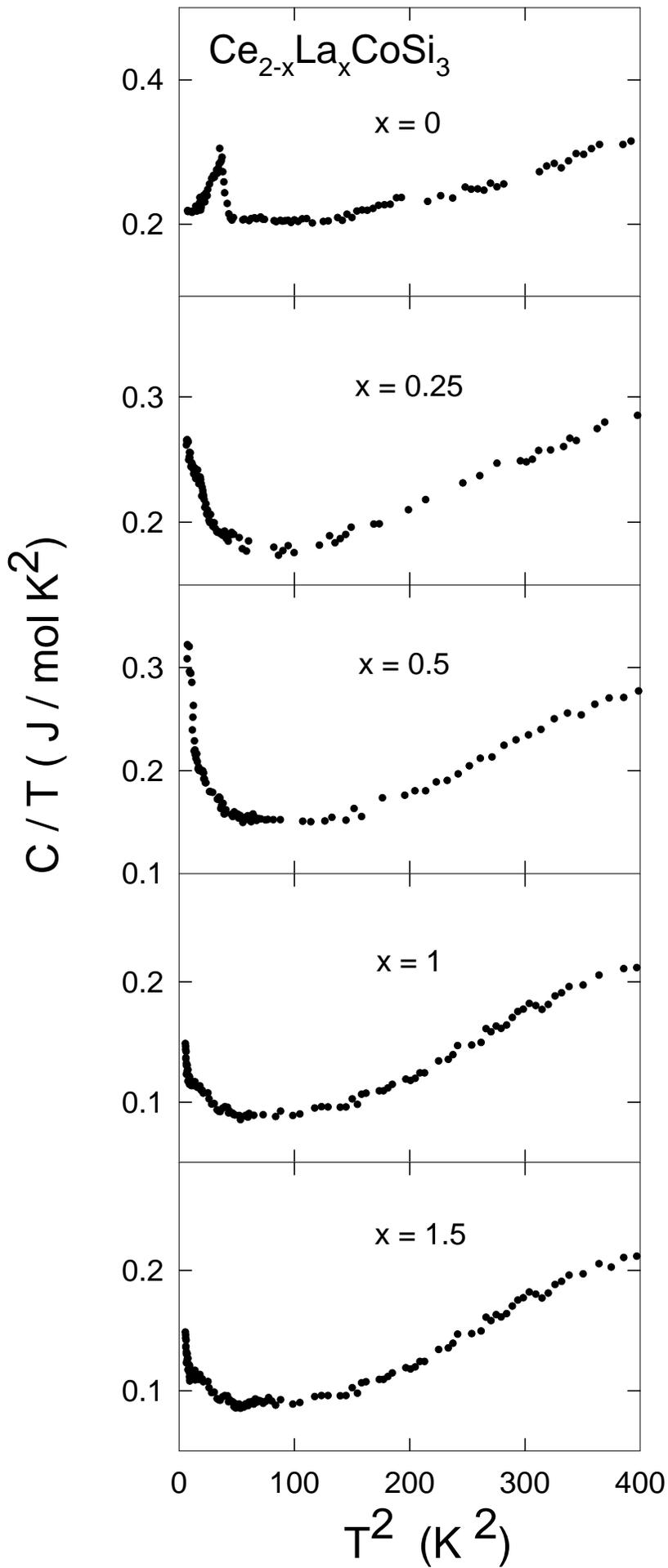
\begin{figure} 
\caption{The heat
capacity divided by temperature versus square of temperature below 20 K
for the alloys, Ce$_{2-x}$La$_x$CoSi$_3$.} 
\end{figure} 
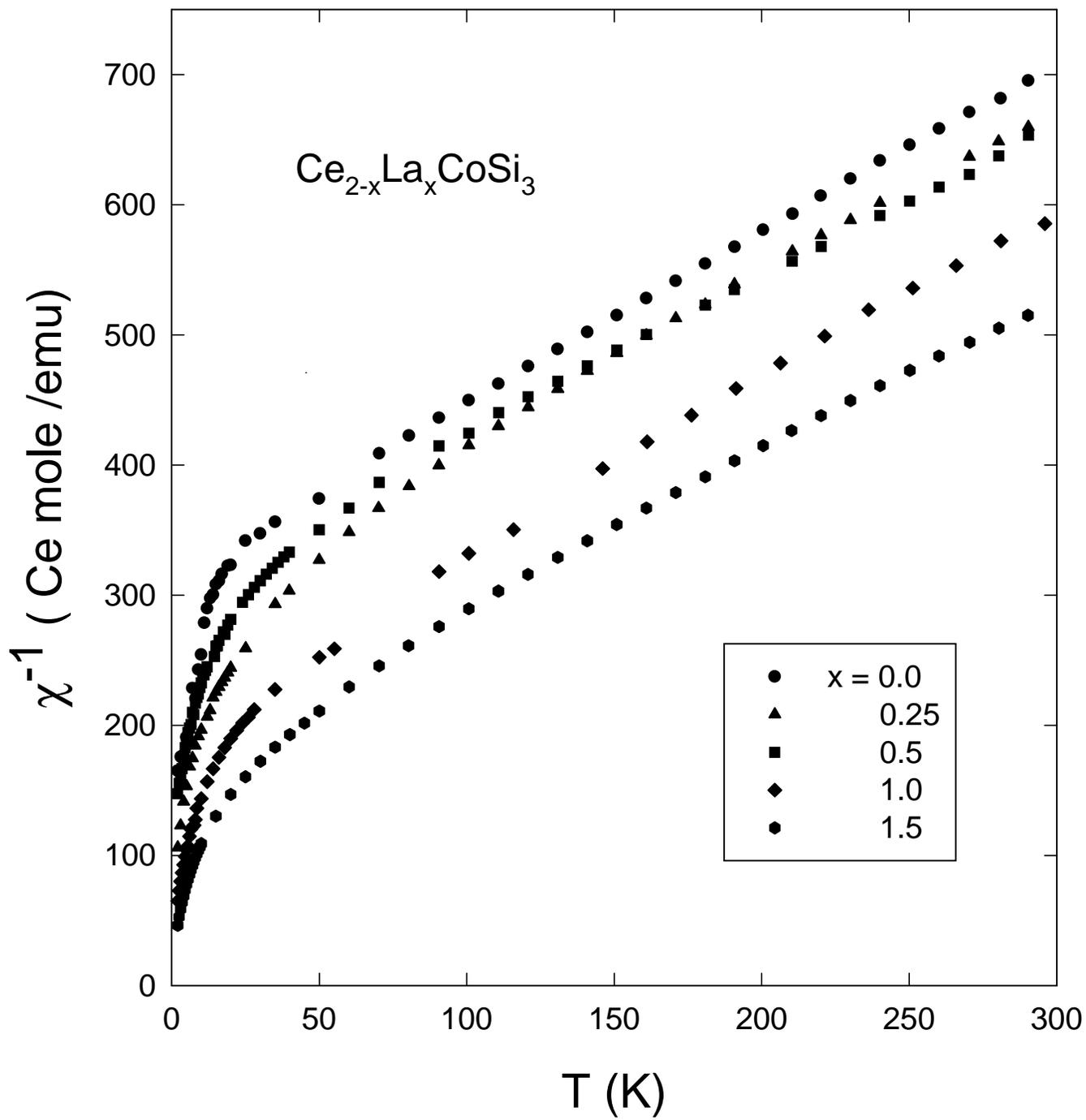
\begin{figure}
\caption{Inverse susceptibility per Ce mole (after subtracting the 300 K
$\chi$ value of La$_2$CoSi$_3$) as a function of temperature, for
Ce$_{2-x}$La$_x$CoSi$_3$.} 
\end{figure} 
\end{document}